\documentclass{article}
\usepackage{fleqn}
\usepackage{graphicx}
\def\bibname{References}%

\def\thebibliography#1{\section*{\bibname\markboth
 {\bibname}{\bibname}}\small \list
 {}{\setlength\labelwidth{1.4em}\leftmargin\labelwidth
 \setlength\parsep{0pt}\setlength\itemsep{0pt}
 \setlength{\itemindent}{-\leftmargin}
 \usecounter{enumi}}
 \def\newblock{\hskip .11em plus .33em minus -.07em}
 \sloppy
 \sfcode`\.=1000\relax}

\begin{document}
\LARGE
\begin{center}
The Psychopathological Fabric of Time (and Space) and Its Underpinning Pencil-Borne Geometries
\end{center}
\vspace*{-.0cm}
\Large
\begin{center}
Metod Saniga$^{1}$ and Rosolino Buccheri$^{2}$

\vspace*{.1cm} \small $^{1}${\it Astronomical Institute, Slovak Academy
of Sciences, 05960 Tatransk\' a Lomnica, Slovak Republic}\\
$^{2}${\it Istituto di Astrofisica Spaziale e Fisica Cosmica, CNR,
Via Ugo La Malfa 153, 90146 Palermo, Italy}
\end{center}

\vspace*{-.4cm}
\noindent
\hrulefill

\vspace*{.1cm}
\small
\noindent
{\bf Abstract}\\
\noindent The paper presents, to our knowledge, a first fairly comprehensive and mathematically well-underpinned classification of the psychopathology
of time (and space). After reviewing the most illustrative first-person accounts of `anomalous/peculiar' experiences of time (and, to a lesser degree, also
space) we introduce and describe in detail their algebraic geometrical model. The model features six qualitatively different types of the internal structure
of time dimension and four types of that of space. As for time, the most pronounced are the ordinary `past-present-future,' `present-only'
(`eternal/everlasting now') and
`no-present' (time `standing still') patterns; the remaining types represent an intriguing superposition of the three. Concerning space, the most elementary
are the ordinary, i.e. `here-and-there,' mode and the `here-only' one (`omnipresence'); the remaining two cases are again a specific mixture of the former
two.  We then show what the admissible combinations of temporal and spatial psycho-patterns are and give a rigorous algebraic geometrical classification
of them. The paper ends with a brief account of important epistemological/ontological questions stemming from the approach.

\vspace*{-.1cm}
\noindent \hrulefill

\vspace*{.3cm} \noindent
\large
{\bf 1. Introduction}

\vspace*{.0cm}
\noindent
\normalsize
Time is undoubtedly one of the deepest mysteries science has ever faced. Indeed, one would hardly find something that is, on the one hand, so
intimately connected with our experience and yet, on the other, so difficult to come to grips with. Nothing, perhaps, can better illustrate this point than a
large group of phenomena that are collectively referred to as the {\it psychopathology} of time, that is, all `anomalous/peculiar' experiences of time as
invariably encountered and reported in various mental psychoses, drug-induced states, deep meditative and mystical states as well as in many other
`altered' states of consciousness (see, e.g., Jaspers, 1923; Minkowski, 1933; Melges, 1982; Hartocollis, 1983; Cutting and Silzer, 1990; Saniga, 2000;
and references
therein). For this peculiar fabric of psychological time comprises, as we shall soon see in more detail, such bizarre,
paradoxical and mind-boggling forms as `eternity, everlasting now,' `arrested/suspended' time, time `going backward,' and even
`disordered/fragmented' time, to mention the most pronounced of them.

Up to date, there exists no acceptable psychological/neurological model capable of properly dealing with these fascinating time constructs and
underpinning any logical classification of them. The reason why this is so rests, in our opinion, upon the following two facts.  First, these extraordinary
experiences of time (and, of course, space as well) are inherently participatory, non-reproducible and subjective and, so, seriously at odds with current
methodologies/paradigms of science, which strives for reproducibility and objectivity. Second, the most pronounced departures from the `consensus'
reality are so foreign to our `waking' mind that their properties defy our common sense logic and cannot be adequately communicated in words; an
interested scholar has
to go through a large number of relevant first-hand accounts/narratives and acquire the ability to read between the lines in order to spot an(y)
underlying conceptual pattern. We are therefore convinced that further progress in our understanding of these phenomena will inevitably entail a
serious shift in the corresponding scientific paradigms to reveal their true epistemological/ontological status and be accompanied by the increasing
use of sufficiently abstract mathematical concepts to properly grasp their qualitative properties.

Our study of psychopathological (space-)times has, from the very beginning, been pursued in accordance with this strategy (Saniga, 1998b; 1999; 2000).
The model discussed in the sequel
thus features not only a fairly high level of abstraction, but it also poses a serious challenge to some generally accepted dogmas in natural sciences.
Formally, it employs advanced geometrical concepts, like a projective space and/or Cremona transformations. Conceptually, it relies on a daring and
far-reaching assumption that the anecdotal, first-person descriptions of extraordinary states of consciousness are {\it on a par with} standard
observational/experimental evidence in natural sciences. It is this `abstract geometrization of the first-person perspective' that gives our approach
a remarkable unifying and predictive power and makes it a very promising conceptual step towards the ultimate unveiling of the riddle of time.
The purpose of the paper is to demonstrate this. The presentation is focussed on conceptual issues rather than mathematical technicalities, the latter
being reduced to the extent that also the reader with a comparatively slight mathematical background can easily follow the main line of
reasoning.

%\vspace*{.5cm} \normalsize \noindent
\newpage
\noindent
\large
{\bf 2. Psychopathological (Space)Times: Most Illustrative Cases}

\vspace*{.15cm}
\noindent
\normalsize
We shall start with a compact, yet comprehensive enough, review of the most distinguished forms of `anomalous' experience of time. This review
is unique in that it consists solely of first-person accounts/narratives, three or four per each mode. We decided for this way of the exposition of the
subject so that even the uninitiated reader can get a fairly clear idea about the nature of experiences involved and realize
the source and character of possible difficulties one is likely to face when attempting to mathematically model these experiences.

\vspace*{.3cm}
\noindent
\large
{\it 2.1. `Eternity,' alias `Eternal/Everlasting Now'}

\vspace*{.15cm}
\noindent
\normalsize
This is perhaps the most pronounced and in the literature best-documented kind of profoundly `distorted' sense of time. It is a sort
of compressing, telescoping of past, present and future into the present moment that is experienced as `eternal/everlasting now.'  One
of the best portrayals of what this experience looks like is found in the following account (Huber, 1955):
\footnote{In this and all the subsequent excerpts/quotations we italicise those parts of the narratives that most directly relate to the topic
of the given subsection.}

\vspace*{.15cm}
\noindent
\small
I woke up in a whole different world in which the puzzle of the world was solved extremely easily in a  form of {\it a different space}. I was amazed at the
wonder of this different space and this amazement  concealed my judgement, {\it this space is totally distinct from the one we all know. It had different
dimensions, everything contained everything else. I was this space and this space was me}. The outer space was part of this space, {\it I was in the outer
space and the outer space was in me}...

Anyway, I didn't experience time, time of the outer space and aeons until the second phase of this dream. In the cosmic flow of time you saw worlds
coming into existence, blooming like flowers, actually existing and then disappearing. It was an endless game. {\it If you looked back into the past, you
saw aeons,  if  you looked  forward into the future there were aeons stretching into the eternity, and this eternity was  contained in the point of the present.
One was situated in a state of being in which the `will-be' and the `vanishing' were already included}, and this `being' was my consciousness. It
contained it all. This `being-contained' was presented very vividly in a geometric way in form of circles of different size which  again were all part of a
unity since all of the circles formed exactly one circle. The biggest circle was part of the smallest one and vice versa. As far as the differences of size
are concerned, I could not give any accurate information later on...

\normalsize
\vspace*{.15cm}
\noindent
This narrative is remarkable in a couple of aspects. Not only does the subject try to understand his uncanny experience of time in terms of a
simple {\it geometrical} model, but he also pays particular attention to the spatial fabric of his extraordinary state, which also differs utterly from what is
regarded as a normal/ordinary perception of space; in fact, the subject finds himself to be one/fused with space!

Another impressive description
of the same kind of psycho-spacetime is taken from Atwater (1988). It is based on one of many author's near-death experiences, which
was also accompanied by a fascinating archetypal imaginery:

\vspace*{.15cm}
\noindent
\small
This time, I moved, not my environment, and I moved rapidly... My speed accelerated until I noticed a wide but thin-edged expanse of bright light ahead,
like a `parting' in space or a `lip,' with a brightness so brilliant it was beyond light yet I could look upon it without pain or discomfort... The closer I came
the larger the parting in space appeared until... I was absorbed by it as if engulfed by a force field...

Further movement on my part ceased because of the shock of what happened next. Before me there loomed two gigantic, impossibly huge masses
spinning at great speed, looking for all the world like cyclones. One was inverted over the other, forming an hourglass shape, but where the spouts
should have touched there was instead incredible rays of power shooting out in all directions... I stared at the spectacle before me in disbelief...

As I stared, I came to recognize my former Phyllis self in the midupperleft of the top cyclone. Even though only a speck, I could see my Phyllis clearly,
and {\it superimposed over her were all her past lives and all her future lives happening at the same time in the same place as her present life. Everything
was happening at once!} Around Phyllis was everyone else she had known and around them many others... The same phenomenon was happening to
each and all. {\it Past, present, and future were not separated but, instead, interpenetrated like a multiple hologram combined with its own reflection.}

The only physical movement anyone or anything made was to contract and expand. {\it There was no up or down, right or left, forward or backward. There
was only in and out,} like breathing, like the universe and all creation were breathing -- inhale/exhale, contraction/expansion, in/out, off/on.

\normalsize
\vspace*{.15cm}

The last, but by no means less astounding than the former two, example of this subsection, borrowed from Braud (1995), depicts in great detail and clarity
a gradual transformation of our ordinary, waking sense of time (and space) into that of `eternity' (and `omnipresence'):

\vspace*{.15cm}
\noindent
\small
...I get up and walk to the kitchen, thinking about what a timeless experience would be like. I direct my attention to everything that is happening at the
present moment -- what is happening here, locally, inside of me and near me, but non-locally as well, at ever increasing distances from me. I am
imagining everything that is going on in a slice of the present -- throughout the country, the planet, the universe. It's all happening at once.

{\it I begin to
collapse time, expanding the slice of the present, filling it with what has occurred in the immediate `past.'} I call my attention to what I just did and
experienced, what led up to this moment, locally, but keep these events within a slowly expanding present moment. {\it The present slice of time slowly
enlarges, encompassing, still holding, what has gone just before, locally, but increasingly non-locally as well}.  By now, I am standing near the kitchen
sink. {\it The present moment continues to grow, expand. Now it expands into the `future' as well. Events are gradually piling up in this increasingly larger
moment. What began as a thin, moving slice of time, is becoming thicker and thicker, increasingly filled with events from the `present,'`past,' and
`future.' The moving window of the present becomes wider and wider, and moves increasingly outwardly in two temporal directions at once.}  It is as
though things are piling up in an ever-widening present.

The `now' is becoming very thick and crowded! {\it `Past' events do not fall away and cease to be;
rather, they continue and occupy this ever-widening present. `Future' events already are, and they, too, are filling this increasingly thick and full present
moment. The moment continues to grow, expand, fill, until it contains all things, all events.  It is so full, so crowded, so thick, that everything begins to
blend together. Distinctions blur. Boundaries melt away. Everything becomes increasingly homogeneous, like an infinite expanse of gelatine. My own
boundaries dissolve. My individuality melts away. The moment is so full that there no longer are separate things.} There is no-thing here. There are no
distinctions.

A very strong emotion overtakes me. Tears of wonder-joy fill my eyes.  This is a profoundly moving experience. Somehow, I have moved
away from the sink and am now several feet away, facing in the opposite direction, standing near the dining room table. I am out of time and in {\it an
eternal present.}  In this present is everything and no-thing. I, myself, am no longer here. Images fade away. Words and thoughts fade away. Awareness
remains, but it is a different sort of awareness. Since distinctions have vanished, there is nothing to know and no one to do the knowing. `I' am no longer
localized, but no longer `conscious' in the usual sense. There is no-thing to be witnessed, and yet there is still a witnesser.

The experience begins to
fade. I am  `myself' again. I am profoundly moved. I feel awe and great gratitude for this experience with which I have been blessed...

\vspace*{.3cm}
\noindent
\large
{\it 2.2. Time `Standing Still,' alias `Arrested/Suspended' Time }

\vspace*{.15cm}
\noindent
\normalsize
Another well-documented and quite abundant anomalous temporal mode. A couple of spectacular examples are found in Tellenbach (1956; p.\,13):

\vspace*{.15cm}
\noindent
\small
I sure do notice the passing of time but couldn't experience it. I know that tomorrow will be another day again but don't feel it approaching. I can estimate
the past in terms of years but I don't have any connection to it anymore. The {\it time standstill is infinite,} I live in a constant eternity. I see the clocks
turn but {\it for me time does not flow}... Everything lies in one line, there are {\it no differences of depth anymore}... Everything is like a firm {\it plane}...

\normalsize
\vspace*{.15cm}
\noindent
and ({\it ibid}; p.\,14):

\vspace*{.15cm}
\noindent
\small
Everything  is very different in my case, time is passing very slowly. Nights last so long, one hour is as long as usually a whole day... Sometimes {\it
time had totally stood still,} it would have been horrifying. Even {\it space had changed}: Everything is so empty and dark, everything is so far away
from me...

{\it I don't see space as usual, I see everything as if it were just a background. It all seems to me like a wall, everything is flat.} Everything
presses down, everything looks away from me and laughs...

\normalsize
\vspace*{.15cm}
\noindent
Both reports are by depressive (melancholic) patients. It is worth noticing here that when time comes to a stillstand, perceived space
seems to lose one dimension, becoming thus two-dimensional.
A slightly more detailed description of this time pattern we succeeded in finding in a nicelly written paper by Muscatello and Giovanardi Rossi
(1967; p.\,784):

\vspace*{.15cm}
\noindent
\small
{\it Time is standing still for me,} I believe. It is perhaps only a few moments that I have been so bad. I look at a clock and I have the impression, if I look
at it again, that an {\it enormous} period of time has passed, as if hours would have passed instead only a few minutes. {\it It seems to me that a duration
of time is enormous. Time does not pass any longer, I look at the clock but its hands are always at the same position, they no longer move, they no longer
go on;  then I check if the clock came to a halt, I see that it works, but the hands are standing still.} I do not think about my past, I remember it but I do not
think about it too much. When I am so bad, I never think about my past. Nothing enters my mind, nothing... I did not manage to think about anything. I did
not manage to see anything in my future. The present does not exist for me when I am so bad... the past does not exist, the future does not exist.

\normalsize
\vspace*{.15cm}
\noindent
The following vignette, taken form a treatise on mescaline-induced experimental psychoses  by Beringer (1969; p.\,311), is also impressive:

\vspace*{.15cm}
\noindent
\small
...The strangest thing was that every once in a while my normal time-awareness, as far as these figures were concerned, got totally lost; {\it time was no
longer a stream, which flew away and whose flux could have been measured, but it was rather similar to a sea, which as a whole stood still and which
was in itself only a chaotic and utter jumble}. I was no longer able to understand the continuous becoming of the figures as a sequence in a certain time
direction, but sometimes the colours and forms flew into an indescribable jumble, as if the previously alternating figures were now experienced all
simultaneously. Had I previously seen these figures in a constant motion, so now it was only a colorful and inexpressible manifold there in which I was
not able to perceive any motion anymore. When I totally sank into the show of the figures, it happened every now and then that {\it I also sank into this
time-still-standing, where the succession was transformed into a still standing present}. Not only am I now not able to formulate these interruptions of
the normal time experience, I am also almost unable to imagine my experience of them any more. When I tore myself away from these figures and
violently turned myself to the outer world, this anomalous time experience was no longer here, but this disturbance of the sense of time found its
expression in a form of illusion that an {\it immense long} time must have passed since my last waking-up.

\vspace*{.3cm}
\noindent
\large
{\it 2.3. Time `Going/Flowing Backward'}

\vspace*{.15cm}
\noindent
\normalsize
This kind of time pathology is very often found in mental psychoses. Here is a very illuminating and particularly representative case,
communicated by a schizophrenic patient
(Fischer, 1929; p.\,556):

\vspace*{.15cm}
\noindent
\small
Yesterday at noon, when the meal was being served, I looked at the clock: why did no one else? But there was something strange about it. For the clock
did not help me any more and did not have anything to say to me any more. How was I going to relate to the clock? {\it I felt as if I had been put back, as if
something of the past returned, so to speak, toward me, as if I were going on a journey. It was as if at 11:30 a.m. it was 11:00 a.m. again, but not only
 time repeated itself again, but all that had happened for me during that time as well.} In fact, all of this is much too profound for me to express. In the
middle of all this something happened which did not seem to belong here. {\it Suddenly, it was not only 11:00 a.m. again, but a time which passed a long
time before was there and there inside} -- have I already told you about a nut in a great, hard shell? It was like that again: {\it in the middle of time I was
coming from the past towards myself.} It was dreadful. I told myself that perhaps the clock had been set back, the orderlies wanted to play a stupid trick
with the clock. I tried to envisage time as usual, but I could not do it; and then came a feeling of horrible expectation that I could be sucked up into the
past, or that the past would overcome me and flow over me. It was disquieting that someone could play with time like that, somewhat daemonic...

\normalsize
\vspace*{.15cm}
\noindent
A brief and concise description of `psycho-time-reversal' is found in Laing (1968):

\vspace*{.15cm}
\noindent
\small
{\it ... I got the impression that time was flowing backward; I felt that time proceeded in the opposite direction,} I had just this extraordinary sensation,
indeed... {\it the most important sensation at that moment was, time in the opposite direction...} The perception was so real that I looked at a clock and,
I do not know how, I had the impression that the clock confirmed this feeling, although I was not able to discern the motion of its hands...

\normalsize
\vspace*{.15cm}
\noindent
A strikingly similar, but much more articulated, depiction is also furnished by a depressive patient of Kloos (1938; p.\,237):

\vspace*{.15cm}
\noindent
\small
As I suddenly broke down I had this feeling inside me that time had completely flown away. After those three weeks in a sick-camp, I had this feeling
that the clock hands run idle, that they do not have any hold. This was my sudden feeling. I did not find, so to speak, any hold of a clock and of life
anymore, I experienced a dreadful psychological breakdown. I do not know the reason why I especially became conscious of the clock. At the same
time, {\it I had this feeling that the clock hands run backward...} There is only one piece left, so to speak, and that stands still. I could not believe that time
really did advance, and that is why I thought that the clock hands did not have any hold and ran idle... As I worked and worked again, and worried and
did not manage anything, {\it I simply had this feeling that everything around us (including us) goes back...} In my sickness I simply did not come along
and  then {\it I had this delusion inside me that time runs backward...} I did not know what was what anymore, and I always thought that I was losing my
mind. I always thought that the clock hands run the wrong way round, that they are without any meaning. I just stood-up in the sick-camp and looked at
the clock -- and it came to me then at once: well, what is this, time runs the wrong way round?!... I saw, of course, that the hands moved forward, but, as I
could not believe it, {\it I kept thinking that in reality the clock runs backward...}

\vspace*{.3cm}
\noindent
\large
{\it 2.4. `Disordered/Fragmented' Time}

\vspace*{.15cm}
\noindent
\normalsize
The following experience, voluntarily induced by mescaline, is the most representative one we have been able to find in the literature available
(Ebin, 1961):

\vspace*{.15cm}
\noindent
\small
For half an hour nothing happened. Then I began feeling sick; and various nerves and muscles started twitching unpleasantly. Then, as this wore off, my
body became more or less anaesthetized, and I became `de-personalized', i.e., I felt completely detached from my body and the world...

This experience alone would have fully justified the entire experiment for me..., but at about 1.30 all interest in these visual phenomena was abruptly
swept aside when I found that {\it time was behaving even more strangely than color.} Though perfectly rational and wide-awake... {\it I was not
experiencing
events in the normal sequence of time. I was experiencing the events of 3.30 before the events of 3.0; the events of 2.0 after the events of 2.45, and
so on. Several events I experienced with an equal degree of reality more than once. I am not suggesting, of course, that the events of 3.30 happened
before the events of 3.0, or that any event happened more than once. All I am saying is that I experienced them, not in the familiar sequence of clock
time, but in a different, apparently capricious sequence which was outside my control.}

By `I' in this context I mean, of course, my disembodied self, and by `experienced' I mean learned by a special kind of awareness which seemed to
comprehend yet be different from seeing, hearing, etc.... I count this experience, which occurred when, as I say, I was wide awake and intelligent, sitting
in my own armchair at home, as the most astounding and thought-provoking of my life...

\normalsize
\vspace*{.15cm}
\noindent
And here is another intriguing mescaline-borne episode of a very similar time's sense (Beringer, 1969; p.\,148):

\vspace*{.15cm}
\noindent
\small
...While walking upstairs, a sudden and as if nailed-down picture of this moment, the momentary view of Dr. M., Dr. St. and myself in space, attracted
my attention. This repeated itself on different stairs. At the top of the stairway {\it there seemed to be no continutity of time at all, the whole course of
events was only a mess of separate situations without any connection.} And these situations, in case of active work, could later have been connected in
the same way in which one can observe a celluloid film. Yet at the same time these situations -- in both experiencing and a direct reproduction of the
happening afterwards -- carried the character of the {\it in}dependent and {\it dis}connected. {\it A strange next-to-each-other-ness, not a
one-after-the-other-ness;} they have no position in time, time has no sense here...

\vspace*{.3cm}
\noindent
\large
{\bf 3. Pencil-Borne Dimensions of Time (and Space) and Psychopathology}

\vspace*{.15cm}
\noindent
\normalsize
From the accounts sampled in the previous section it is quite obvious that the fabric of psychological time is so intricate, complex and
multifarious that, at first sight, it may seem to lie completely beyond grasp of any mathematical framework.  Yet, the contrary is true.  In what
follows we shall introduce and describe in detail a simple algebraic geometrical model that not only is capable of qualitatively accounting for all
the `peculiar' time structures mentioned above, but also predicts some novel forms of these.

\begin{figure}[t]
\centerline{\includegraphics[width=14.0truecm,clip=]{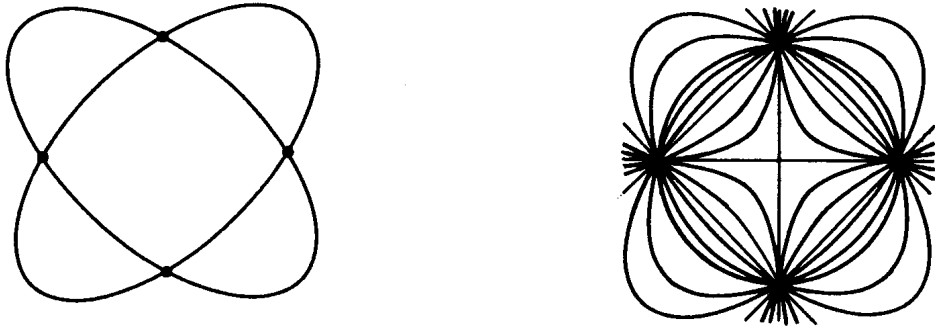}}
\caption{Two distinct conics (ellipses) in the real plane ({\it left}) define a unique pencil ({\it right}). This pencil, which is the most general one,
features four simple base points and three distinct composite conics of the same type (a pair of distinct lines). }
\end{figure}

\vspace*{.3cm}
\noindent
\large
{\it 3.1. Time Dimension as a Pencil of Conics and its (Psychopathological) Patterns}

\vspace*{.1cm}
\noindent
\normalsize
A cornerstone of our model are {\it conics}, in particular their simplest, i.e. linear and single-parametric, aggregates, usually called {\it pencils} (see, e.g.,
Levy, 1964; Saniga, 1998a). Here, linearity means that only two distinct conics are needed to define a pencil (see Fig.\,1), whereas single-parametricity
signifies that the aggregate is a one-dimensional geometrical structure. A point of intersection of two conics is clearly incident with all conics of their
pencil; this point is called a {\it base} point of the pencil. A pencil of conics features, as easily discernible from Fig.\,1, up to four base points. These, of
course, need not be all distinct. It then follows that there exist different kinds of a pencil. In the case of four base points, we find as many as five
different types, namely: all the four points simple; one double point and two simple points; two double points; one triple and one simple point; and
a single point of multiplicity four. A conic is analytically defined by a second order (quadratic) equation and it is {\it composite} (singular) or {\it
proper} according as the equation is factorable or not. A hyperbola, a parabola and an ellipse are all (and the only) examples of proper conics;
a composite conic consists of either a pair of lines, which can be distinct or coinciding, or a single point. Any pencil of conics in the {\it real} plane
contains at least one composite conic, and maximum three (not necessarily distinct and/or of the same type). Given a pencil of conics, a point
of the plane that is not a base point of the pencil is incident with {\it exactly one} (possibly composite) conic of the pencil. Any other proper conic of the
pencil then necessarily belongs into one of two qualitatively different, disjoint families according as the point (henceforth referred to as the reference
point) lies in its {\it in}terior or {\it ex}terior;
\footnote{Given a proper conic, a point, not on the conic, is its exterior or interior point depending on whether or not it lies on a line tangent
to the conic. The exterior/interior of the conic is the set of all its exterior/interior points.}
in what follows we shall call such proper conics, respectively, {\it in}-conics or {\it ex}-conics, and the unique conic incident with the reference
point, if being proper, will be denoted as the {\it on}-conic.

At this point it should already be obvious that it is this a-pencil-of-conics-and-a-point configuration that is of greatest interest and importance to us,
because, in the case where the reference point falls on a proper conic (see Fig.\,2a), it lends itself as a natural, elegant and remarkably simple
explanation of our ordinary experience/sense of time. The only assumption to be made to see this is, following Saniga (2003), to conceive
{\it each proper conic} of the pencil
as a {\it single event/moment} of time, with the understanding that the set of {\it ex}-conics represent events of the {\it past}, that of {\it in}-conics
stand for events of
the {\it future} and the unique {\it on}-conic generates the {\it present} moment, the {\it `now.'}

\begin{figure}[t]
\centerline{\includegraphics[width=\textwidth,height=5.0truecm,clip=]{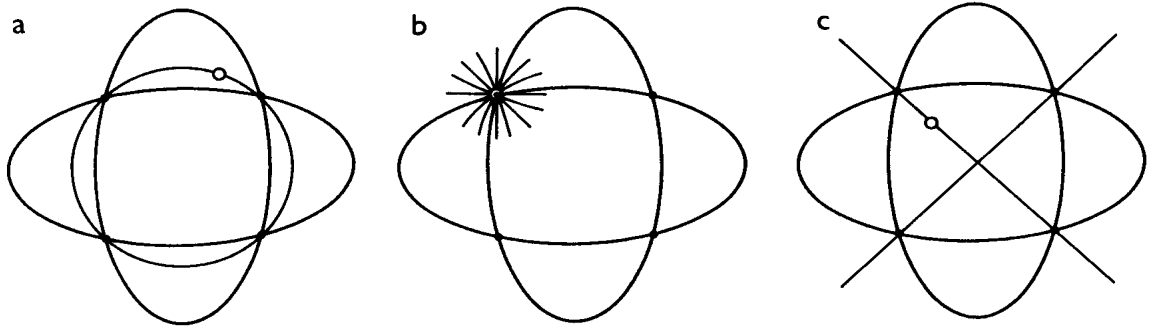}}
\caption{The three qualitatively different patterns of the pencil-borne temporal dimension according as the reference point (a small circle)
is ({\it a}) incident with a proper conic, ({\it b}) coincides with one of the base points, or ({\it c}) falls on a composite conic of the generating
pencil.  The pencil is, as in Fig.\,1,  of the most general type. }
\end{figure}

Apart from this noteworthy `past-present-future' pattern, mimicking everyone's common sense of time, our model gives rise to other two prominent,
in a sense dual to each other, structures.
These correspond, as the attentive reader may have already noticed,
to the cases where the reference point coincides with a base point of the pencil (Fig.\,2b), or falls on one of its composite conics
(Fig.\,2c). In the former case, clearly, all the proper conics are on-conics, whereas in the latter case the pattern is lacking any such conic, being endowed
with ex- and in-conics only. Hence, the corresponding time dimension, in the former case, consists solely of the present moments (the `present-only'
mode), whilst, in the latter case, it comprises only the past and future, being devoid of the moment of the present (the `no-present' mode).  Now, it is
the right point to return to the previous section and try to rephrase  these two unusual temporal arrangements in terms of pathological temporal
constructs listed there.
In doing so, we readily find out that the {\it present-only} pattern nicely accounts for nothing but experiences of {\it `eternity,'  `everlasting now'}
(Sect.\,2.1). Concerning the {\it no-present} design, here a little thought suggests that this is a proper fit for the time {\it `standing still'} mode (Sect.\,2.2);
for our feeling that time `flows,' `proceeds' is unequivocally tied to the notion of the present moment, the `now,' as the linking element between the
past and future and so it is only too natural to assume that the absence of this element in the pattern should correspond to a complete
suspension/cessation of the (sense of) time's flow.

Before we proceed to examine other intriguing structures of pencil-time, and attempt geometrization of the remaining two queer forms of
time's experiences dealt with in Section 2, it is instructive to make a slight digression  to discuss a very interesting feature of our approach that
has a serious bearing on the very meaning of the term `pathological' when it comes to the concept of time. This feature tells us about a relative
`probability' of the occurrence of the above-discussed three patterns of time in the realm of psychopathology. This probability should not, however,
be understood in a strict sense of the word, but rather in a looser, algebraic geometrical sense. The reasoning goes as follows. The conics of any
pencil sweep up the whole plane and as the latter contains $\infty^{2}$ (double infinity) of points, there are $\infty^{2}$ of potential past-present-future
patterns. Next, as our pencil features three composite conics, each of these is a pair of distinct lines, and a line possesses $\infty^{1}$ (single
infinity) of points, we have $3 \times 2 \times \infty^{1} = 6 \times \infty^{1} \approx \infty^{1}$ of no-present modes. And, finally, as our pencil
features four base points, there are just four present-only structures. We see a clear predominance of the past-present-future mode within the group;
no wonder that it corresponds to our `ordinary,' `consensus' experience of time. Equivalently, this explains why experiences of `eternity' and/or time
`standing still' are regarded/referred to as `anomalous/peculiar;' for the relative probability of their occurrence with respect to our `ordinary' experience
of time is truly negligible.

\begin{figure}[t]
\centerline{\includegraphics[width=\textwidth, height=5.0truecm,clip=]{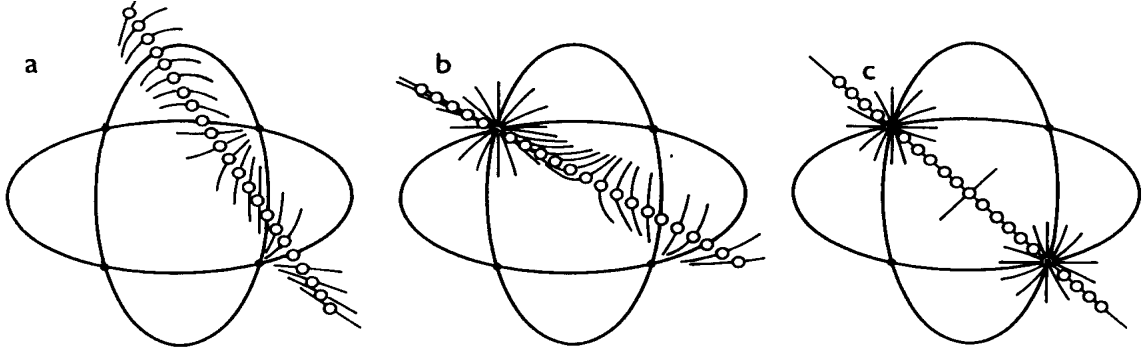}}
\caption{The three distinct types of `line-related' pencil-patterns of time in dependence on whether the reference line is incident with ({\it a}) zero,
({\it b}) one, or ({\it c}) two base points of the pencil. In each case, every illustrated point (a small circle) of the line is accompanied by a
drawing of a small portion of the conic incident with this particular point.}
\end{figure}

Let us next turn, as promised, to examine other conceivable forms of our generic pencil-time. We shall assume that instead of a single reference point
there is a whole infinity of them, and these are, for simplicity, taken to form a line. What different kinds of time dimension do we find in this
case? Remarkably, there are, like in the previous case, three of them. They differ from each other, as depicted in Fig.\,3, in the position of this
line with respect to the base points of the generating pencil of conics, being in the sequel labelled, respectively, as a zero-, one- and
two-point pattern according as the reference line hits no, one or two of the base points. Obviously, these line-related temporal structures can each be
regarded as composed of an infinite number of basic, point-related patterns. This composition reads:

\vspace*{.2cm}
\begin{center}
\begin{tabular}{||c||c|c|c||}
\hline \hline type of pattern & past-present-future & present-only & no-present \\
\hline \hline zero-point  & infinity & none & six \\
\hline one-point & infinity & one & three  \\
\hline two-point & none & two & infinity  \\
\hline \hline
\end{tabular}
\end{center}

\vspace*{.2cm}
\noindent
The numbers in the first two columns are readily discernible from Fig.\,3 and the definition of the corresponding elementary modes. It is only the last
(no-present) column that requires a word of explanation. Thus, the number in the first line (six) is the number of intersections of the reference line with
the composite conics of the pencil; it follows from the facts that our pencil features three composite conics, each of these is a pair of distinct lines, and in
a projective plane every line is incident with any other line. The number in the second line (three) answers to the fact that if the reference line passes
through a single base point, the latter absorbs three of these six points. Finally, when the reference line joins two base points, it becomes
a component of a composite conic, i.e. every point of it lies on the composite conic in question. We further see that, among the composites, only one,
the one-point mode (Fig.\,3b), features all the three types of elementary patterns, and, similarly, only one, the two-point mode (Fig.\,3c), lacks the most
familiar of them. On the other hand, there is only one elementary pattern, the no-present one, that enters all the three kinds of composites, and only one,
the present-only mode, whose number is always finite. It is very intriguing to see that there is no homogeneous, i.e. featuring just one elementary
pattern, composite.

What are the phenomenological counterparts of these composite temporal patterns? Clearly, each of them must be a mixture/superposition of the
time's experiences we have found to correspond to the elementary patterns involved. And these are strange constructs indeed. Thus, the
zero-point mode corresponds to such an uncanny state of consciousness where the subject encounters an infinite tangle of `ordinary' experiences
of time, differing from each other in the location of the moment of the present and, consequently, in the spans of the regions of past/future, this
perception being accompanied by the sense of time `standing still.' The one-point case is even weirder, as it includes, on top of the above, also
the feeling of `eternity.'  And these experiences are very much like those of `disordered/fragmented' time given in Sect.\,2.4 (but see also the last
account of Sect.\,2.2)! But what about the two-point structure, an intricate blend of the sense of `eternity/everlasting now' and that of time `standing still'?
This kind of experience was privately communicated to one of us by Linda Howe, an instructor in the `akashic records' (Howe, 1999):

\vspace*{.15cm}
\noindent
\small
One common scenario is when the sense of the self is so expanded, beyond any physical boundary... In this aspect, the awareness of being one
with, or a part of, all that is can be profound. The illusion of separation can be perceived as dissolving and, at the same time, the awareness of the
oneness, or unity,... becomes heightened, sometimes acutely so...

In this the experience of time is dramatic in its expansion and simultaneous contraction. There is a sense that {\it there is only one moment, that
all of time/eternity is held in that instant}, very compressed and as powerful as one's imagination can conceive. {\it Simultaneously}, there is a
sense that {\it there is no time} in the expansion. That {\it all is holding still}. Not even slow motion, but no motion. A {\it total suspension of time} is
experienced. This is
the {\it all time/no time} paradox.

\normalsize
\vspace*{.15cm}
At this point, we have already succeeded in elucidating three out of the four principal types of the pathology of time's sense enumerated in the 
previous section. The only  mode that is left to be explained is the experience/sense of time `going/flowing backward' (Sect.\,2.3). To this end, we 
shall return to our ordinary, past-present-future pattern (Fig.\,2a) and examine what happens to this pattern as the reference point starts 
`moving' away from its original position.  This `motion,' as delineated in Fig.\,4a-c, is assumed to take place in such a way that the point always 
remains incident with one and the same conic. As it is quite obvious from this figure, the qualitative structure of the original pattern (Fig.\,4a) is 
preserved until the reference point, en route, hits a base point (Fig.\,4b), in which case the pattern acquires its present-only type. Further motion 
of the reference point clearly leads to re-establishment of the original type, but with one remarkable difference -- with the in-conics and ex-conics 
having {\it swapped} their roles (Fig.\,4c)! This means nothing but that the time's arrows generated by the two past-present-future patterns,  
although {\it sharing}
the same present moment, point in the {\it opposite} directions! One could hardly find a more elementary explanation of time-reversal...
  
\begin{figure}[t]
\centerline{\includegraphics[width=\textwidth,height=4.5truecm,clip=]{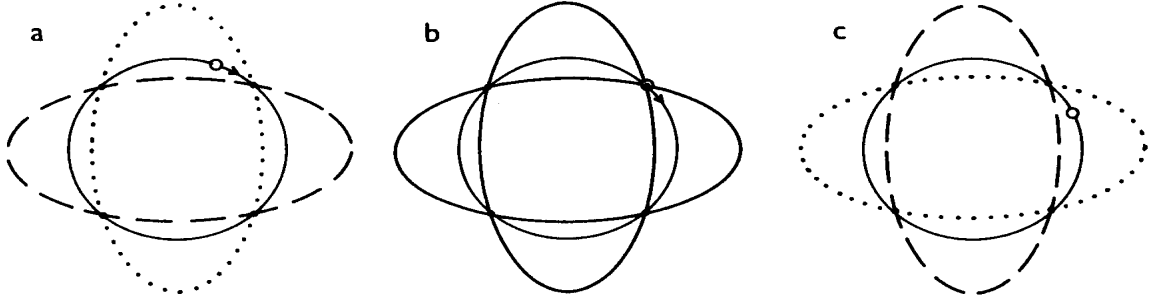}}
\caption{An elementary  explanation of the phenomenon of a psychological time-reversal in terms of our pencil-borne model of time dimension. 
The heavy curve(s) is (are) the on-conic(s), while those drawn as dotted/dashed represent the in-/out-conics. A little arrow indicates the direction 
of the motion of the reference point (a small circle).}
\end{figure}

\vspace*{.3cm}
\noindent
\large
{\it 3.2. Space Dimension as a Pencil of Lines and its (Psychopathological) Patterns}

\vspace*{.1cm}
\noindent
\normalsize
It is evident that that the concept of a pencil, with conics as its constituting elements, turns out to be an extremely fertile framework for getting a
deeper qualitative insight into the fine structure of psychological time dimension. Motivated by this finding, it is only too natural to address also
the structure of psychological space in a similar  fashion, i.e. retaining the concept, replacing only its constituting elements. As to our senses
space appears to have a less complex structure than time, and a line is a simpler geometrical object than a proper conic, we shall take a {\it
spatial} dimension to be represented by a pencil of {\it lines} (Saniga, 1998a; 2003).  Our reasoning will parallel that of the previous (sub)section,
which will enable us to see how our approach gets to grips with the fundamental difference between time and space at the perceptual level.

\begin{figure}[t]
\centerline{\includegraphics[width=12.0truecm,height=3.7truecm,clip=]{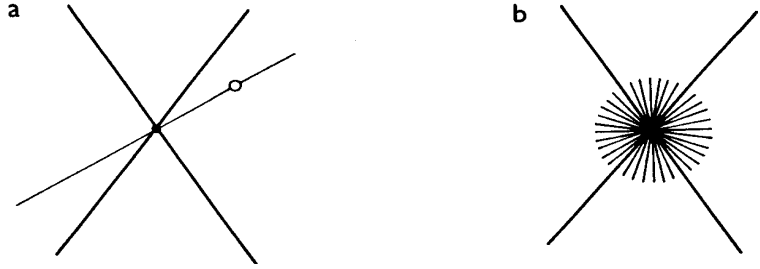}}
\caption{The two qualitatively different elementary patterns of the pencil-borne space dimension depending on whether the reference point is different
({\it a}) or not ({\it b}) from the vertex of the generating pencil of lines. Compare with Figs.\,2{\it a} and 2{\it b}, respectively.}
\end{figure}

\begin{figure}[b]
\centerline{\includegraphics[width=12.0truecm,height=3.7truecm,clip=]{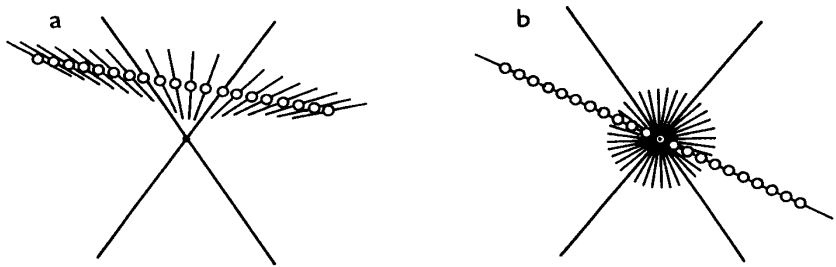}}
\caption{The two composite modes of space pencil-borne dimension, differing from each other in the position of the reference line (illustrated as a range
of small circles) with respect to the vertex of the generating pencil of lines. Similarly to Fig.\,3, every illustrated reference point goes with a drawing of a
small part of the line incident with this particular point.}
\end{figure}

Two distinct lines in a plane suffice to define a unique pencil, i.e. the set of all lines of the plane that pass through the point shared by the two (the latter
being called the {\it vertex} of the pencil). As any two lines in a projective plane have always one, and only one, point in common, there exists only one
type of a pencil of lines; this is the first fundamental difference from the case of conics.  Given a pencil of lines, a point of the plane (the reference point)
that is different from the vertex of the pencil is incident with {\it exactly one} line of the pencil (Fig.\,5a); this line will henceforth be called the {\it
on}-line, the remaining lines of the pencil being termed {\it off}-lines. This particular a-pencil-of-lines-and-a-point configuration qualitatively mimics our
`ordinary,' `here-and-there' sense of space, with the {\it on}-line standing for {\it `here'} and {\it off}-lines for {\it `there.'} It is a spatial
counterpart of the `ordinary,' past-present-future pattern of time (Fig.\,2a). However, it must be pointed out here that, unlike off-conics, off-lines have
all {\it the same} footing with respect to the reference point; this feature thus serves as a nice explanation why, in our `ordinary' state of consciousness,
perceived space has a rather trivial structure when compared to that of perceived time. Another point-related spatial pattern is the one where the
reference point is identical with the vertex of the pencil (Fig.\,5b); as now all the lines of the pencils are on-lines, we get the `here-only' structure.
Being a twin of the `eternity,' `everlasting now' mode (Fig.\,2b), this structure must necessarily be inherent to those `non-ordinary' states of consciousness
that are characterized by feelings of `omnipresence,' or `fusion/oneness' with the universe (see the first and last accounts in Sect. 2.1, as well as
the account in Sect.\,3.1). These here-and-there and here-only modes are obviously the only elementary patterns of pencil-space; for a line is so simple an
object that there exist no singular forms of it and, so, there does not exist any spatial analogue of the no-present pattern. And as there are $\infty^{2}$
potential here-and-there modes, but just a single here-only one, it is only natural that it is the former that underlies our `consensus' perception of space.

The cases with the reference line are also structurally simpler than those of time dimension. There are, as the analogy suggests, a couple of them
according as the line avoids the vertex (Fig.\,6a) or is incident with it (Fig.\,6b).  As it can easily be recognized from Fig.\,6, the former case is a compound
of a single infinity of sole here-and-there modes, whereas the latter features a combination of both the elementary modes, with the preponderance of
the more familiar of them. Accordingly, a subject experiencing the `avoiding-vertex' mode feels to be localized at every point (`multipresent')
along the particular space dimension, while that in a state backed by the `hitting-vertex' mode should feel to be both localized at a particular position of
and simultaneously stretched out along the dimension in question.

\vspace*{.3cm}
\noindent
\large
{\bf 4. Pencil-Borne Space-Time and the Varieties of its Internal Structure}

\vspace*{.15cm}
\noindent
\normalsize
So far we have treated time and space as two completely unrelated dimensions, which is of course in marked contrast to how the two
aspects of reality are perceived to exist. Moreover, we have dealt with a single space dimension only, while our senses tell us that
there are (at least) three of them. So we have accordingly to refine our model to comply with these `constraints.'

To furnish this task, it is necessary to move from the (projective) plane into the (projective) space and -- following and extending of what we did in
Saniga (2001; 2003) -- consider a specific geometrical configuration comprising {\it three} distinct, non-coplanar pencils of lines (generating spatial
dimensions) and a {\it single} pencil of conics (time). The planes carrying the pencils of lines are taken to be collinear, i.e. having a line in common, and
none of the vertices of the pencils (denoted as B$_{i}$, $i$=$1, 2, 3$) is assumed to lie on this shared line (${\cal L}^{{\rm B}}$).  The pencil of conics is,
naturally, situated in the plane defined by the three vertices, and its base points are these vertices and the point (L) of incidence of the plane and the line
${\cal L}^{{\rm B}}$, as portrayed in Fig.\,7.\footnote{In a {\it projective} space, every line is incident with every plane, and also every two planes have a
line in common.} The uninitiated reader may get an impression that our option for this configuration is completely arbitrary. This is, however, not the
case, for this configuration plays a prominent role in the theory of so-called Cremona transformations between two projective spaces of dimension three.
A proper explanation of what a Cremona transformation is and what kind(s) of distinguished structures it entails would, however, take us too far afield
from the main topic of this paper: the interested reader is therefore referred to consult our above-mentioned papers (Saniga, 2001; 2003) and/or a -- though
for first reading a bit difficult -- book by Hudson (1927). For what follows it suffices that the reader shares our intuitive belief that there is indeed
something special to the above-described four-pencil configuration so that Nature found it worth making use of.

\begin{figure}[t]
\centerline{\includegraphics[width=8.0truecm,clip=]{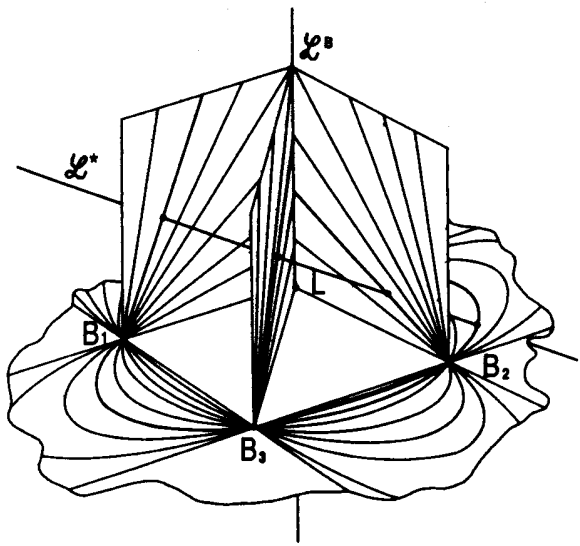}}
\caption{A particular geometrical configuration comprising three pencils of lines and a single pencil of conics, the latter being located in the
plane defined by the vertices of the pencils of lines. The symbols are explained in the text.}
\end{figure}

It is evident that this remarkable configuration, as it stands, can represent only a bare space-time, i.e. the space-time devoid of any subject/observer.
So, in order to introduce the latter into our model, the configuration has to be endowed with an additional geometrical object. This can, of course, be done 
in a number of ways, one of the simplest being in terms of a single line (denoted as ${\cal L}^{*}$ in Fig.\,7). Armed with this premise and the postulates 
and findings of the previous section, we are able to find out which kinds of spatial and temporal patterns discussed above are mutually compatible (i.e., 
can form and `live together' on a single manifold) and thereby arrive at a first fairly comprehensive and mathematically well-underpinned classification 
of the psychopathology of time and space. It is not hard to see that this task simply boils down to examining all possible positions of the reference line 
${\cal L}^{*}$ within this configuration that lead to qualitatively different arrangements of pencil-patterns induced by the point(s) of intersection of the line 
with the four pencil-carrying planes.

We shall, of course, start with the case when the reference line is in a generic position with respect to the four planes. As it is obvious from Fig.\,7, in this 
case the line cuts each of these planes in a unique point. As this point is clearly different from any of the three vertices and from the point L as well, it 
specifies in each of the three planes B$_{i}{\cal L}^{{\rm B}}$ (henceforth simply $l$-planes) a unique line, and in the B$_{1}$B$_{2}$B$_{3}$-plane 
($c$-plane) a unique, in general proper, conic. Each of the three pencils of lines thus generates the here-and-there mode, and the pencil of conics features 
the past-present-future pattern. So, our {\it generic} pencil-borne space-time is, as expected, the space-time as perceived in our {\it ordinary} state of 
consciousness.

In order to facilitate our subsequent discussion, we shall compactify our notation for different kinds of pencil-borne patterns. For each elementary
pattern we shall reserve one letter, uppercase for time and lowercase for space; a composite mode will then bear several letters, corresponding 
to the elementary modes it consists of. As for time pencil-patterns, we shall adopt the following symbols: `A' for the ordinary, past-present-future
mode (the `arrow' of time); `E' for the present-only (`eternity') mode; and `S' for the no-present (time `standing still') mode. The composite modes
will then have the following abbreviations: `$\mu$A$\cdot$S' for the zero-point mode; `$\mu$A$\cdot$S$\cdot$E' for the one-point mode; and 
`S$\cdot$E' for the two-point
mode, with $\mu$ standing for `multi-'. Concerning space patterns, we shall use `h' for the ordinary, here-and-there mode and `o' for the here-only 
(`omnipresence') mode. Its composites will accordingly be denoted as `$\mu$h' (vertex-avoiding) and `h$\cdot$o' (vertex-hitting).

So what are possible kinds of our pencil-borne, `Cremonian' psycho-space-times? From an algebraic geometrical point of view, there are altogether 
19 different types of them, as depicted in Fig.\,8. And they are seen to form a truly remarkable sequence, once being grouped into distinict
rows according to their number/abundance (the first column) and into distinct columns according to the number of dimensions of localizability 
(the lowermost row) and/or the character of the Cremonian image of the reference line in the second projective space (the uppermost row). As easily 
discernible from the figure, the individual sub-figures differ from each other in the position of the reference line and each of them is accompanied by 
four of the above-introduced labels/acronyms so that the reader can readily find out the corresponding internal pattern of each spatial dimension and 
time as well. The number/abundance of a particular type within the structure is, as above, of a geometrical origin. Thus, there are (see, e.g., Sommerville, 
1951) $\infty^{4}$ (quadruple infinity) of lines in a three dimensional projective space and out of them $\infty^{3}$ are incident with a given line, 
$\infty^{2}$ with two different (possibly incident) lines, and $\infty^{1}$ pass through a given point and simultaneously lie in a given plane; a line 
is uniquely defined by two distinct points (their joint) or two different planes (their meet). 
Non-localizability in a particular dimension means that the reference line does {\it not} define a unique line in the corresponding $l$-plane, or
a unique conic in the $c$-plane. Hence,  `o', `$\mu$h' and `h$\cdot$o'  are non-local patterns of a space dimension, while `E', `$\mu$A$\cdot$S', 
`$\mu$A$\cdot$S$\cdot$E' and `S$\cdot$E' are those of time; in Fig.\,8, the former/latter are illustrated by drawing several lines/conics in the 
corresponding $l$-planes/$c$-plane so that they can readily be recognized.

Just a passing look at the figure in question reveals a number of intriguing facts. First, and perhaps the most crucial one, is that our consensus 
space-time (represented by the sub-figure in the top left-hand corner; this sub-figure is a fully equivalent  version of Fig.\,7) is, as expected, by far the 
{\it most} abundant type in the hierarchy, 
as there are $\infty^{4}$ of its potential cases. On the other hand, there is {\it just one} potential case of space-time where the subject is {\it completely} 
non-localized (the sub-figure in the bottom right-hand corner; the reference line is here identical with the line ${\cal L}^{{\rm B}}$).  Next, it is fairly 
obvious 
that the most numerous patterns are those where the subject is completely localized (the `all' column); as the number of dimensions of non-localizability
{\it in}creases (i.e., as we move in the figure from left to right), the number of potential cases {\it de}creases (i.e., we move from the top to the bottom of
the figure). 
Further, we notice that if there are
at least two dimensions of non-localizability (the last three columns), one of them is always time. It is also intriguing to observe that 
as for two non-ordinary elementary patterns of time, the S-mode prevails over the E-mode. The most variegated row is seen to be the $\infty^{2}$-one 
(featuring six different types of space-time patterns and spanning three different levels of non-localizability), the least variegated
being the top and bottom ones. It is also worth stressing that out of the spatial modes it is the h-one that occurs most frequently, while
amongst the temporal patterns it is the S- and E-modes that enjoy this property. Also, there exists no pencil-borne space-time whose space dimensions 
would be all of the o- or h$\cdot$o-type. Interestingly, the least frequently encountered patterns are the h$\cdot$o (space) and 
S$\cdot$E (time) ones. Finally, there are pairs of patterns which are incompatible with each other: the o-mode with the $\mu$h-one, the 
S$\cdot$E-mode with the $\mu$h-one and the S-pattern with the o-one.

\begin{figure}[t]
\centerline{\includegraphics[width=\textwidth,height=10.0truecm,clip=]{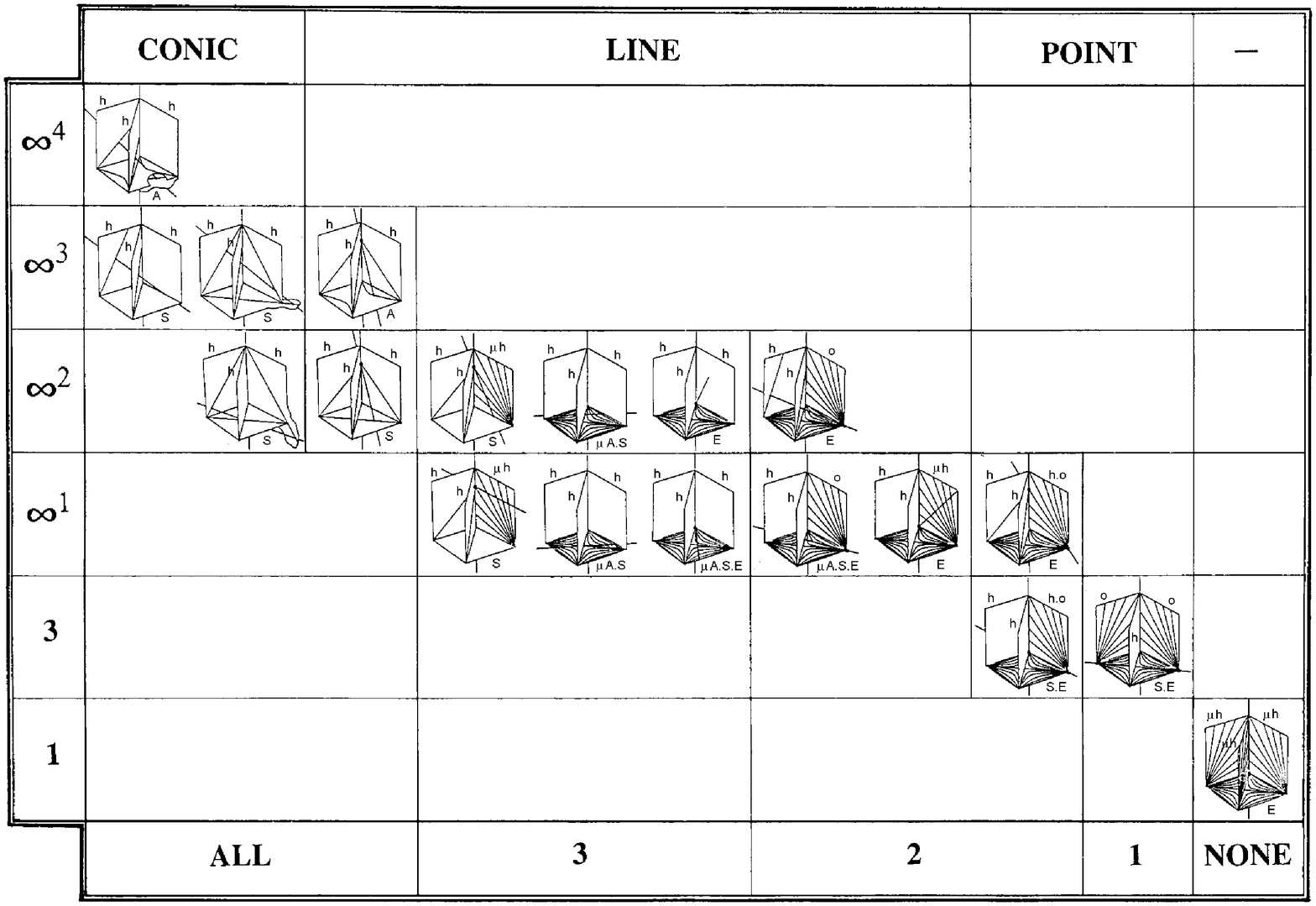}}
\caption{A diagrammatic sketch of an algebraic geometrical classification of pencil-borne space-times. All the symbols and notation are explained in the 
text.}
\end{figure}

From the information gathered in Fig.\,8 and the findings of the previous section(s) it will represent no difficulty for the reader to infer
and analyze the `experiential contents' for each type of space-time. We only add the following note. One of the most distinguished features of
a great majority of extraordinary states of consciousness is a seriously altered sense of individuality, ego, or self-hood. In particular, the greater 
the departure from our consensus reality, the lesser the sense of ego; ultimately, in the most abstract states, the subject feels to completely 
transcend/surpass his/her sense of ego, and, so,  the dichotomy between subject and object  as well (see the last account in Sect.\,2.1).
This important feature, too, has a proper place in our model, once we identify the `degree' of the sense of ego with the level of localization
of the subject in our pencil-borne space-times. From Fig.\,8 we then readily discern that our consensus experience of space-time
is characterized by the strongest sense of the self. As we move across the figure from left to right, the sense of ego accompanying the 
individual types of space-times (or, better, the corresponding states of consciousness) gradually `melts/dissolves,' until it completely vanishes 
in the state represented by the sub-figure located at the very bottom of the figure. Here is a recently found first-person account (Anonym, 
2003) that describes in great detail not only this transformation of the sense of `I', but also accompanying profound changes in the perception 
of both time and space, and which dovetails very nicely with the implications of our model:

\vspace*{.15cm}
\noindent
\small
\dots For twelve hours I moved in and out of dimensions of both space and time. The incomprehensible became comprehensible. Realities
    within realities blossomed and faded. From the infinitely large to the infinitely small, unbounded and unfettered mind flashed across
    landscapes of incredible depth and beauty\dots
I was looking into the source of my very being, and without question, my creator. And then I came
    to realize too that I was at the interface between individual mind and absolute mind.

    Entheogens, or in my case psilocybes, provide the pivotal role of interfacing between individual consciousness and universal
    consciousness. It is the crucial link or conduit that bridge the two at a single point. That point then begins to widen, and both entities
    slowly merge. As the interface grows, what were initially two now opens into one. It's not just a random happening though, an alignment
    process between the two takes place. Actually it's more a matter of one aligning itself to the other. This is not a conscious operation,
    although consciousness is witness to it\dots
    To experience this phase of the psychic event was an absolute revelation
    with all the glory and beauty imaginable. With my minds eye I was able to see the outline of the interface where the two became the
    One, where duality merged into unity\dots I had the pleasant ability at the center of the interface to
    merge in and out at will. {\it In one moment I was myself, a separate thinking entity with all my individual thoughts; as I merged out my
    self-hood ceased to exist; my individuality gone; my thoughts as unique things ceased to be, given way to absolute thought. Time and
    space played an interesting part in this experience. While in myself time existed, time flowed, there was past and future, but while
    merged in unity time ceased, there was no past or future. Everything was in a single instant; what Plotinus called the `Eternal Now.' In
    myself space had dimension, there was up and down, limitations existed. Merged in the other, there was no up, no down, no limitation,
    all was infinite and absolute.} This gave rise to another incredible phenomenon; with time suspended and space without boundary
    omniscience came into full awareness; yes, all things known; no limitations to knowledge\dots
    Omnipotence, and omnipresence also became an awesome recognition, {\it but not related to me personally since the I had
    ceased to be;} they were aspects of that great Oneness that was the universe of consciousness. Merging back into my own ego left me
    with only a memory of being present to it all\dots

\vspace*{.3cm}
\noindent
\large
{\bf 5. Conclusion}

\vspace*{.15cm}
\noindent
\normalsize
A few weeks before his death, in a letter of condolence to the family of his life-time friend Michele Besso, Albert Einstein wrote (Dukas and Hoffmann, 
1972):  `For us believing physicists the distinction between past, present, and future is only an illusion, even if a stubborn one.' We have, however,
rigorously demonstrated that this `illusion' and its most pronounced `peculiarities/anomalies' are underlaid by a definite algebraic geometrical pattern. 
Does it mean that our math is a sort of illusion, too? Or, rather, is it physics that falls short of grasping the true nature of time? There is ever growing
evidence for the latter to be the case (see, e.g., Elitzur and Dolev, 2003). But be it this way or that,
cracking this dilemma will certainly be not only a great leap forward in our knowledge of time, but it will also entail radical shifts in many 
(currently regarded as well-established) paradigms of natural sciences. And we are convinced that our algebraic geometrical approach
will be recognized as an important step towards furnishing this exciting, though formidable, task.

\vspace*{.3cm} 
\noindent
\large
{\bf Acknowledgements}

\vspace*{.15cm}
\noindent
\small
We are very grateful to Mr. Pavol Bend\' {\i}k for painstaking drawing of the figures and Dr. Richard Kom\v z\' {\i}k for the help with conversion of the 
latter into postscript form. We would also like to express our cordial thanks to Miss Daniela Veverkov\' a and Mr. Peter Hahman for translating into 
English all the excerpts taken from journals written in German. This work was supported in part by the 2001/2002 NATO/FNRS Advanced Research 
Fellowship `Algebraic Geometrical Structure of Spacetime,' the 2000--2002 NATO Collaborative Linkage Grant PST.CLG.976850 `Structure of Time and 
Quantum Computing' and the joint 2001--2003 research project between the Italian Research Council and the Slovak Academy of Sciences `The 
Subjective Time and Its Underlying Mathematical Structure.'


\vspace*{-.1cm} \small
\begin{thebibliography}{10}
\bibitem{} Anonym (2003) Mystical mushrooms, an anonymous account posted at http://www.shroomery.org/in- dex/par/22809.

%\vspace*{-.2cm}
\bibitem{} Atwater, P.M.H. (1988) {\it Coming Back to Life. The After-Effects of the Near-Death Experience}, Dodd,
       Mead and Company, New York, Chapter 2.

%\vspace*{-.15cm}
\bibitem{}
Beringer, K. (1969), {\it Der Meskalinrausch, Seine Geschichte und Erscheinungsweise},
Springer-Verlag, Berlin--Heidelberg--New York.

%\vspace*{-.15cm}
\bibitem{}
Braud, W.G. (1995) An experience of timelessness, {\it Exceptional Human Experience} {\bf 13}(1), 64--66.


%\vspace*{-.15cm}
\bibitem{}
Cutting, J. and Silzer, H. (1990) Psychopathology of time in brain disease and schizophrenia, {\it Behav. Neurol.}
{\bf 3}, 197--215.

\bibitem{}
Dukas, H. and Hoffmann, B. (1972) {\it Albert Einstein: Creator and Rebel}, Hart-Davis and MacGibbon, London, p.\,258. 

\bibitem{}
Ebin, D. (ed.) (1961) {\it The Drug Experience}, Orion Press, New York, p. 295.

\bibitem{}
Elitzur, A.C. and Dolev, S. (2003) Is there more to t? Why time's description in modern physics is still incomplete, in R. Buccheri, M. Saniga and 
W.M. Stuckey (eds.), {\it  The Nature of Time: Geometry, Physics and Perception} (NATO ARW), Kluwer Academic Publishers, Dordrecht--Boston--London, 
pp.\,297--306.

%\vspace*{-.15cm}
\bibitem{}
Fischer, F. (1929) Zeitstruktur und Schizophrenie, {\it Zeitschr. ges. Neurol. Psychiat.} {\bf 121}, 544--574.

%\vspace*{-.15cm}
\bibitem{} Hartocollis, P. (1983) {\it Time and Timelessness, or the Varieties of Temporal Experience}, International Universities
Press, New York.

\bibitem{} Howe, L. (1999) {\it Time and space relationships in non-ordinary states of consciousness}, private communication.

\bibitem{} Huber, G. (1955) {\it Akasa, der Mystische Raum}, Origo-Verlag, Z\" urich, pp.\,45--46.

\bibitem{}Hudson H.P. (1927) {\it Cremona transformations in plane and space},
Cambridge University Press, Cambridge.

%\vspace*{-.15cm}
\bibitem{} Jaspers, K. (1923) {\it Allgemeine Psychopathologie}, Springer-Verlag, Berlin.

%\vspace*{-.15cm}
\bibitem{}
Kloos, G. (1938) St\" orungen des Zeiterlebens in der endogenen Depression, {\it Nervenarzt} {\bf 11}, 225--244.

%\vspace*{-.15cm}
\bibitem{}
Laing, R.D. (1968) {\it La Politica dell'Esperienza}, Feltrinelli, Milano, p. 148.

%\vspace*{-.15cm}

\bibitem{}
Levy, H. (1964) {\it Projective and Related Geometries}, The Macmillan Company, New York, pp.\,251--269.

\bibitem{} Melges, E.T. (1982) {\it Time and the Inner Future: A Temporal Approach to Psychiatric Disorders}, John Wiley and
Sons, New York.

%\vspace*{-.15cm}
\bibitem{} Minkowski, E. (1933) {\it Le Temps V\' ecu -- \' Etudes Ph\' enom\' enologiques et Psychopathologiques}, Gauthier, Paris.

%\vspace*{-.15cm}
\bibitem{}
Muscatello, C.F. and Giovanardi Rossi, P. (1967) Perdita della visione mentale e patologia
       dell'esperienza temporale, {\it Giorn. Psichiatr. Neuropatol.} {\bf 95}, 765--788.

%\vspace*{-.15cm}
\bibitem{}
Saniga, M. (1998a) Pencils of conics: a means towards a deeper understanding of the arrow of time?, {\it Chaos, Solitons $\&$ Fractals} {\bf 9}, 1071-1086.

\bibitem{} Saniga, M. (1998b) Unveiling the nature of time: altered states of consciousness and pencil-generated space-times, {\it
International Journal of Transdisciplinary Studies} {\bf 2}, 8--17.

%\vspace*{-.15cm}
\bibitem{} Saniga, M. (1999) Geometry of psycho(patho)logical space-times: a clue to resolving the enigma of time?, {\it Noetic Journal}
{\bf 2}, 265--273.

%\vspace*{-.15cm}
\bibitem{} Saniga, M. (2000) Algebraic geometry: a tool for resolving the enigma of time?, in R. Buccheri, V. Di Ges\` u and M. Saniga (eds.),
{\it Studies on the Structure of Time: From Physics to Psycho(patho)logy}, Kluwer Academic/Plenum Publishers, New York, pp.\,137--166.

\bibitem{} Saniga, M. (2001) Cremona transformations and the conundrum of dimensionality and signature of macro-spacetime, {\it Chaos, Solitons
$\&$ Fractals} {\bf 12}, 2127--2142.

\bibitem{} Saniga, M. (2003) Geometry of time and dimensionality of space, in R. Buccheri, M. Saniga and W.M. Stuckey (eds.), {\it  The
Nature of Time: Geometry, Physics and Perception} (NATO ARW), Kluwer Academic Publishers, Dordrecht--Boston--London, pp.\,131--143;
also physics/0301003.

\bibitem{} Sommerville, D.M.Y. (1951) {\it Analytical Geometry of Three Dimensions}, Cambridge University Press, Cambridge.

%\vspace*{-.15cm}
\bibitem{}
Tellenbach, H. (1956) Die Raumlichkeit der Melancholischen. I. Mitteilung, {\it Nervenarzt} {\bf 27}, 12--18.


%\vspace*{-.15cm}
%\bibitem{} Saniga M. Cremona transformations and the conundrum of
%dimensionality and signature of macro-spacetime. Chaos, Solitons
%$\&$ Fractals 2001;12:2127--42.

%\vspace*{-.15cm}
%\bibitem{} Saniga M. On `spatially anisotropic' pencil-space-times associated
%with a quadro-cubic Cremona  transformation. Chaos, Solitons $\&$
%Fractals 2002;13:807--14.

%\vspace*{-.15cm}
%\bibitem{} Saniga M. Quadro-quartic Cremona transformations and
%four-dimensional pencil-space-times with the reverse signature. Chaos,
%Solitons $\&$ Fractals 2002;13:797--805.

\end{thebibliography}
\end{document}